\documentclass[a4paper,10pt,twoside]{cpc-hepnp}

\usepackage{multicol}
\usepackage{graphicx}
\usepackage{booktabs}
\usepackage{amssymb,bm,mathrsfs,bbm,amscd}
\usepackage[tbtags]{amsmath}
\usepackage{lastpage}

\begin{document}

\fancyhead[c]{\small Submitted to ¡®Chinese Physics C '} 


\title{Reexamination of constrains on the Maxwell-Boltzmann distribution by Helioseismology\thanks{Supported by National Natural Science Foundation of China (11135005, 11021504) and the Major State Basic Research Development Program of China (2013CB834406)}}

\author{%
HE Jian-jun$^{1;1)}$\email{jianjunhe@impcas.ac.cn}%
\quad ZHANG Li-yong$^{1,2,3}$%
\quad HOU Su-qing$^{1,3}$
\quad XU Shi-wei$^{1,3}$\email{shwxu@impcas.ac.cn}
}
\maketitle

\address{%
$^1$ Institute of Modern Physics, Chinese Academy of Sciences, Lanzhou 730000, China\\
$^2$ School of Nuclear Science and Technology, Lanzhou University, Lanzhou 730000, China\\
$^3$ University of Chinese Academy of Sciences, Beijing 100049, China\\
}

\begin{abstract}
Nuclear reactions in stars occur between nuclei in the high-energy tail of the energy distribution and are sensitive to possible
deviations from the standard equilibrium thermal-energy distribution, the well-known Maxwell-Boltzmann Distribution (\textsf{MBD}).
In a previous paper published in Physics Letters 441B(1998)291, Degl'Innocenti {\it et al}. made strong constrains on such deviations
with the detailed helioseismic information of the solar structure. With a small deviation parameterized with a factor
exp$[{-\delta (E/kT)^2}]$, it was shown $\delta$ restricted between -0.005 and +0.002. These constrains have been carefully reexamined
in the present work. We find that a normalization factor was missed in the previous modified \textsf{MBD}. In this work, the
normalization factor $c$ is calculated as a function of $\delta$. It shows the factor $c$ is almost unity within the range
0$< \delta \leq$0.002, which supports the previous conclusion. However, it demonstrates that $\delta$ cannot take a negative value
from the normalization point of view. As a result, a stronger constraint on $\delta$ is defined as 0$\leq \delta \leq$0.002. The
astrophysical implication on the solar neutrino fluxes is simply discussed based on a positive $\delta$ value of 0.003. The reduction
of the $^7$Be and $^8$B neutrino fluxes expected from the modified \textsf{MBD} can possibly shed alternative light on the solar
neutrino problem. In addition, the resonant reaction rates for the $^{14}$N($p$,$\gamma$)$^{15}$O reaction are calculated with a
standard \textsf{MBD} and a modified \textsf{MBD}, respectively. It shows that the rates are quite sensitive even to a very small
$\delta$. This work demonstrates the importance and necessity of experimental verification or test of the well-known \textsf{MBD} at
high temperatures.
\end{abstract}



\footnotetext[0]{\hspace*{-3mm}\raisebox{0.3ex}{$\scriptstyle\copyright$}2013
Chinese Physical Society and the Institute of High Energy Physics
of the Chinese Academy of Sciences and the Institute
of Modern Physics of the Chinese Academy of Sciences and IOP Publishing Ltd}%

\begin{multicols}{2}

\section{Introduction}
Under the ideal condition of non-interaction states, infinite volume and zero density, a single scale (the temperature or the average
one-body energy) characterizes all the equilibrium distributions, which are described by the Maxwell-Boltzmann distribution
(\textsf{MBD})~\cite{bib:deg98,bib:lan80,bib:huang}. However, it is well-known that the actual distribution, which deviates from the
standard \textsf{MBD}, is characterized by additional scales (total energy, Fermi energy, \textit{etc}.)~\cite{bib:deg98}. The ideal
\textsf{MBD} has been widely accepted and utilized in the nuclear astrophysical community~\cite{bib:fow67,bib:cla83,bib:rol88},
thereinto, the MBD for the single particle energy distribution can be written as
\begin{eqnarray}
f_\mathrm{MBD}(E)=\frac{2}{\sqrt{\pi}}\frac{\sqrt{E}}{(kT)^{3/2}}e^{-E/kT}
\label{eq:one}.
\end{eqnarray}

In the past, Degl'Innocenti \textit{et al.}~\cite{bib:deg98} made a strong constraint on possible deviations from the standard
\textsf{MBD} based on the theoretical and observational knowledge of solar physics, the so called Helioseismology. In their work,
small deviations of the modified \textsf{MBD} for the collision-energy distribution of the reacting nuclei \textit{i} and \textit{j},
was parameterized to first approximation by introducing a dimensionless parameter $\delta$, and was expressed as (Equ. 4 in
Ref.~\cite{bib:deg98})
\begin{eqnarray}
f_\mathrm{ij}^\delta (E)=f_\mathrm{\textsf{MBD}}(E)e^{-\delta (E/kT)^2}
\label{eq:two}.
\end{eqnarray}
They found that $\delta$ should lie between -0.005 and +0.002 by analyzing the detailed helioseismic information of the solar
structure with a stellar evolutionary code \texttt{FRANCE}~\cite{bib:deg97}. It was shown that even value of $\delta$ as small as
0.003 could give important effects on the solar neutrino fluxes.
In this work, we will make an even stronger constraint on $\delta$ by carefully reexamining the previous work.

\section{Reexamination}
In this Report, we will make a stronger constraint on $\delta$ by taking the normalization issue which was missed in the previous
work into account. Actually, the standard \textsf{MBD} (Equ.~\ref{eq:one}) is normalized to unity within the energy range from 0 to
$\infty$. But the modified \textsf{MBD} (Equ.~\ref{eq:two}) was not normalized to unity in the previous work~\cite{bib:deg98}.
The above Equ.~\ref{eq:two} can be normalized to unity by introducing a factor $c(\delta)$ (a function of $\delta$),
then Equ.~\ref{eq:two} should be read as
\begin{eqnarray}
f_\mathrm{ij}^\delta (E)=c(\delta) \times f_\mathrm{\textsf{MBD}}(E)e^{-\delta (E/kT)^2}
\label{eq:three}.
\end{eqnarray}
The normalization factor $c(\delta)$ can be calculated by
\begin{eqnarray}
c(\delta)=\frac{\sqrt{\pi}}{2}\frac{1}{\int_0^\infty \sqrt{x} \mathrm{exp}(-x-\delta x^2)\mathrm{d}x}
\label{eq:four}.
\end{eqnarray}
Here, $c$($\delta$=0)=1.0 becomes the standard \textsf{MBD} condition.
For the positive $\delta$ values (a depleted high-energy tail), the factor $c(\delta)$ is numerically calculated and shown in
Fig.~\ref{fig1}. It shows that the factor $c$ is quite close to unity (about 1\% error) in the range 0$\leq \delta \leq$0.002
constrained in the previous work~\cite{bib:deg98}. Thus, the previous conclusions are still hold for the positive $\delta$.
However, the integration in Equ.~\ref{eq:four} has no solution (mathematically unconverged) for the negative $\delta$ values
(an enhanced high-energy tail). It means that $\delta$ cannot take a negative value.
Although this convergence issue was 'mathematically' solved by the cut-off technique~\cite{bib:deg98} for the negative $\delta$ values,
this kind of cut-off is actually incorrect in physics. Therefore, a much stronger constraint has to put
on $\delta$ following the previous results (Equ. 16 in Ref.~\cite{bib:deg98}),\textit{ i.e}., 0$\leq \delta \leq$0.002.
Back to Fig. 1 in Ref.~\cite{bib:deg98}, it is shown now three ratios (Y$_\mathrm{ph}$/Y$_\mathrm{ph}^\mathrm{SSM}$,
$\rho_\mathrm{b}$/$\rho_\mathrm{b}^\mathrm{SSM}$ and R$_\mathrm{b}$/R$_\mathrm{b}^\mathrm{SSM}$) can be fitted pretty well for the
positive $\delta$ values. For the largest $\delta$=0.007 data point, small deviations appear in all three ratios, which could be
possibly explained by adding the small factor $c$($\delta$=0.007) (2\%) which was neglected in the previous calculation.
However, reproduction of Fig. 1 in Ref.~\cite{bib:deg98} is beyond the scope of this work.

\begin{center}
\includegraphics[width=6.2cm]{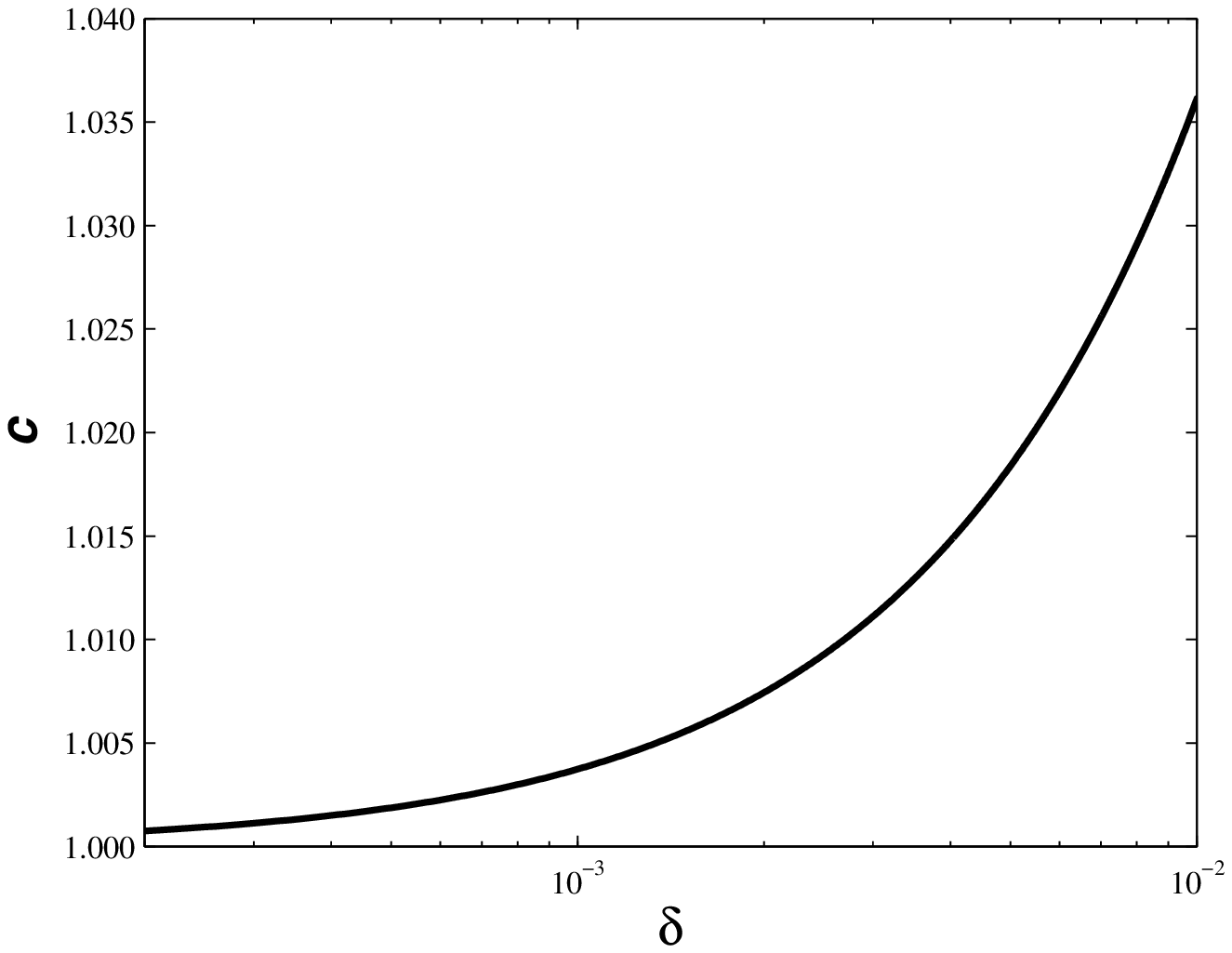}
\figcaption{\label{fig1} Normalization factor $c$ of the modified \textsf{MBD} as a function of $\delta$.}
\end{center}

\begin{center}
\includegraphics[width=6.2cm]{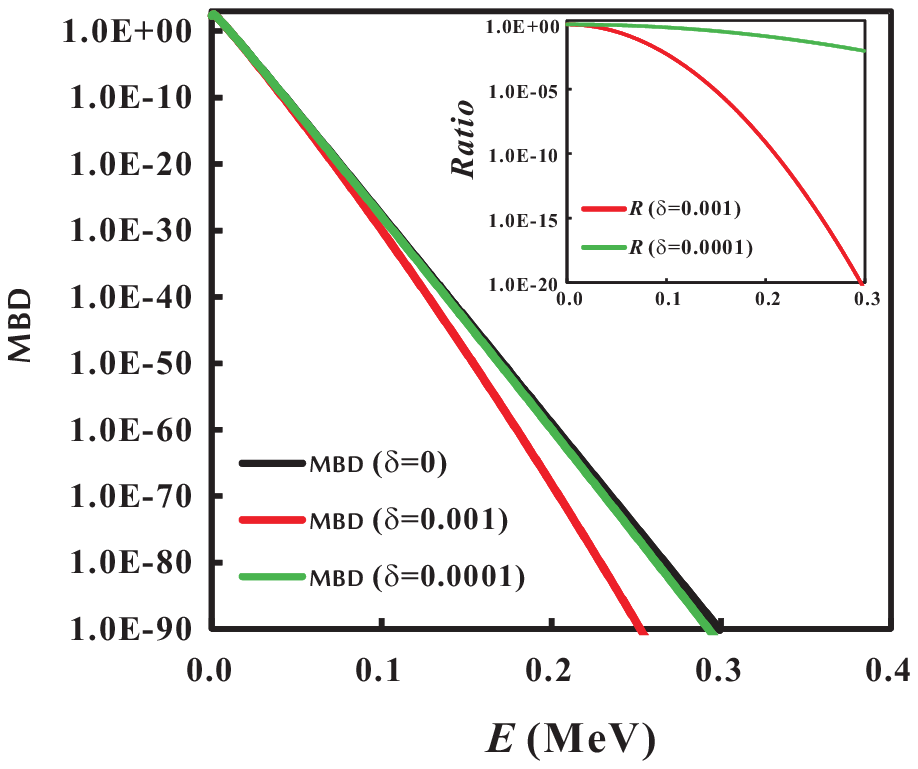}
\figcaption{\label{fig2} (Color online) The standard \textsf{MBD} (Equ.~\ref{eq:one}) and the modified \textsf{MBD} (Equ.~\ref{eq:three}
with $\delta$=0.001, 0.0001) are shown for the core temperature ($T_9$=0.016) of our Sun. The ratios are inserted for comparison.}
\end{center}

\end{multicols}

\ruleup
\begin{center}
\includegraphics[width=10cm]{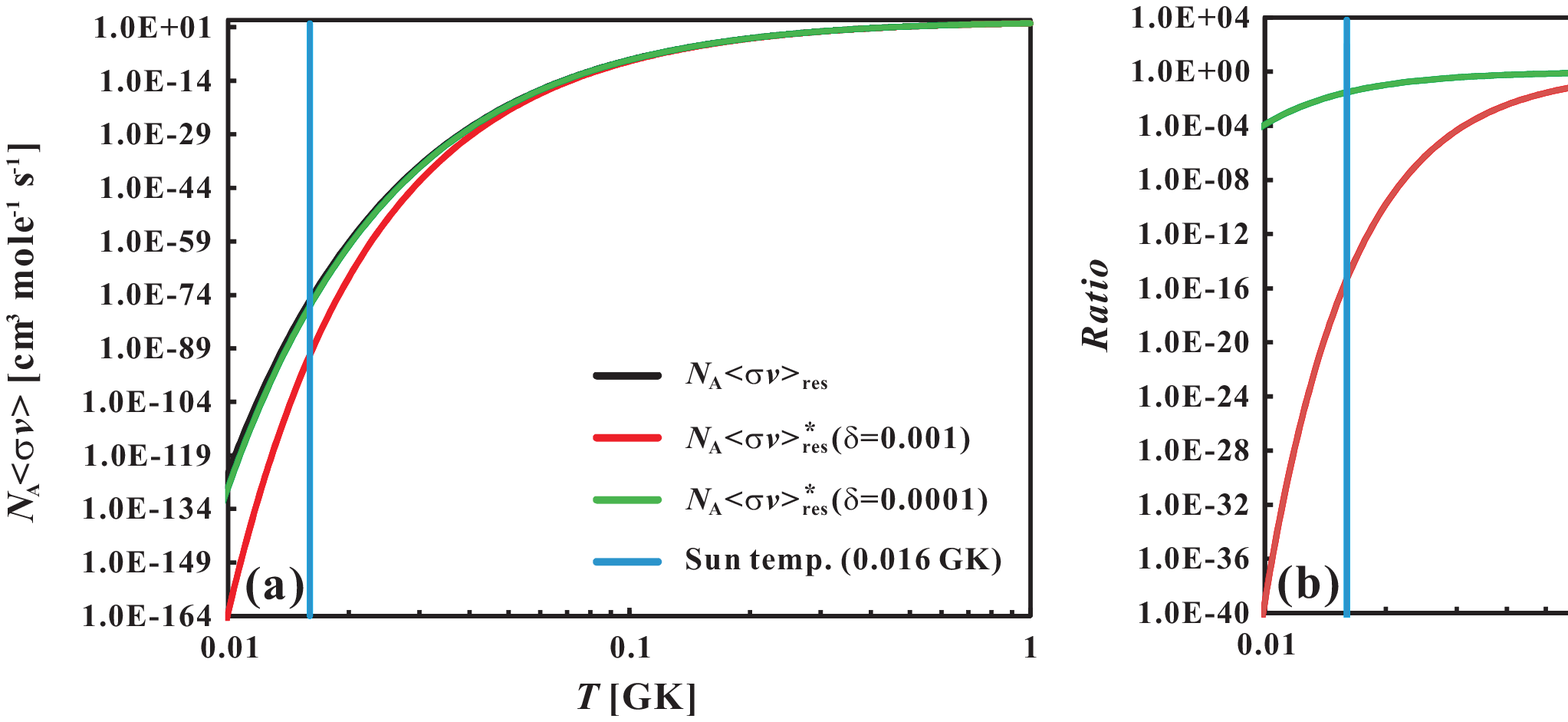}
\figcaption{\label{fig3} (Color online) Resonant reaction rates of the $^{14}$N($p$,$\gamma$)$^{15}$O reaction for the resonance at
$E_r$=0.259 MeV. (a) shows the rates calculated by the standard \textsf{MBD} (in black) and the modified \textsf{MBD} ($\delta$=0.001
in red, $\delta$=0.0001 in green), and (b) shows the corresponding ratios. See text for details.}
\end{center}

\begin{multicols}{2}
Degl'Innocenti \textit{et al.}~\cite{bib:deg98} calculated the effect of nonstandard statistics on the solar neutrino fluxes, and
shew that even for small value of $\delta$ the Boron and Beryllium neutrino fluxes change substantially. For instance, the relative
deviations from Standard Solar Models (\textsf{SSMs}) of the $^7$Be and $^8$B neutrino fluxes are,
$\frac{\Delta \Phi_\mathrm{Be}}{\Phi_\mathrm{Be}}$=-0.30 and $\frac{\Delta \Phi_\mathrm{B}}{\Phi_\mathrm{B}}$=-0.55, respectively,
for $\delta$=+0.003. These pronounced deviations imply that the $^7$Be and $^8$B neutrino fluxes estimated by the modified
\textsf{MBD} are expected to be lower by 30\% and 55\% comparing to those derived from the standard \textsf{MBD} utilized in the
\textsf{SSMs}. These large deviation could possibly shed alternative light on the famous solar neutrino problem~\cite{bib:bah92}.

\section{Application}
The standard \textsf{MBD} and modified \textsf{MBD} as a function of energy are calculated for the core temperature ($T_9$=0.016)
of our Sun as shown in Fig.~\ref{fig2}, and the corresponding ratios are also shown in a small figure inserted. It shows that the
modified \textsf{MBD} is much smaller than the standard one at high-energy tail even for a small $\delta$ value of 0.0001.
This depleted tail will dramatically change the reaction rate, which is of great nuclear astrophysical interests.
For example, the well-known narrow resonance at $E_r$=0.259 MeV ($J^{\pi}$=1/2$^+$, $\omega \gamma$=1.4$\times 10^{-8}$ MeV) in the
compound $^{15}$O nucleus~\cite{bib:ang99} dominates the reaction rate of the $^{14}$N($p$,$\gamma$)$^{15}$O reaction, which rate
governs the efficiency of the \texttt{CNO} cycle, at $T_9 \leq$1. Its resonant reaction rates, according to the analytic
narrow-resonance equation, can be calculated as~\cite{bib:rol88}
\begin{eqnarray}
N_A \langle \sigma v \rangle_\mathrm{res}=1.54 \times 10^{-11} \frac{\omega\gamma}{\mu T_9^{3/2}} \mathrm{exp} \left (-\frac{11.605E_r}{T_9} \right)
\label{eq:five},
\end{eqnarray}
where resonant energy ($E_r$) and strength ($\omega \gamma$) are in units of MeV, and the reduced mass $\mu$ in amu. If the deviation
from the standard \textsf{MBD} is taken into account, the above Equ.~\ref{eq:five} reads as

\end{multicols}
\ruleup
\begin{eqnarray}
N_A \langle \sigma v \rangle_\mathrm{res}^{\ast}=1.54 \times 10^{-11} \frac{\omega\gamma}{\mu T_9^{3/2}} \mathrm{exp} \left [-\frac{11.605E_r}{T_9} - \delta \left (\frac{11.605E_r}{T_9}\right)^2 \right]
\label{eq:six}.
\end{eqnarray}

\begin{multicols}{2}
The rates calculated by Equ.~\ref{eq:five} \&~\ref{eq:six} are plotted in Fig.~\ref{fig3}(a), where $\delta$=0.001, 0.0001 are used
in the calculation. The corresponding ratios
$R=\frac{N_A \langle \sigma v \rangle_\mathrm{res}^{\ast}}{N_A \langle \sigma v \rangle_\mathrm{res}}$ are shown in Fig.~\ref{fig3}(b).
It shows that the rates with the modified \textsf{MBD} are much smaller than those with the standard \textsf{MBD}. Even for a very
small $\delta$=0.0001, the former rate is only about 3\% of the standard value at a typical temperature of $T_9$=0.016 in the core
of our Sun.

\section{Conclusion}
As a conclusion, a small deviation of the \textsf{MBD} can dramatically affect the nuclear reaction rates, which is a very crucial
input parameters for the astrophysical nucleosynthesis models. Therefore, a detailed experimental verification or test of the
well-known \textsf{MBD} at high temperatures is extremely important and necessary for the nuclear astrophysical community.

\vspace{2mm}
\centerline{\rule{80mm}{0.1pt}}

\end{multicols}



\vspace{2mm}

\begin{thebibliography}{90}

\vspace{3mm}
\bibitem{bib:deg98}
Degl'Innocenti S et al. Phys. Lett. B, 1998, \textbf{441}: 291--298
\bibitem{bib:lan80}
Landau L D, Lifshitz E M. Statistical Physics. Oxford: Pergamon Press, 1980
\bibitem{bib:huang}
Huang K. Statistical Mechanics (2nd Edition). New York: John Wiley \& Sons, Inc., 1987, (1987)
\bibitem{bib:rol88}
Rolfs C E, Rodney W S. Cauldrons in the Cosmos. Chicago: Chicago University Press, 1988
\bibitem{bib:fow67}
Fowler W A, Caughlan G R, Zimmerman B A. Ann. Rev. Astron. Astrophys., 1967 \textbf{5}: 525--570
\bibitem{bib:cla83}
Clayton D D. Principles of Stellar Evolution and Nucleosynthesis. Chicago: Chicago University Press, 1983
\bibitem{bib:deg97}
Ciacio F, Degl'Innocenti S, Ricci B. Astr. Astrophy. Suppl. Ser., 1997, \textbf{123}: 449--454
\bibitem{bib:bah92}
Bahcall J N, Pinsonneault M. Rev. Mod. Phys., 1992, \textbf{64}: 885--926
\bibitem{bib:ang99}
Angulo C et al. Nucl. Phys. A, 1999, \textbf{565}: 3--183

\end{thebibliography}
\end{document}